\documentclass[slac_one]{revtex4}
\usepackage{color}
\usepackage{epsfig}
\usepackage{graphicx}
\usepackage{rotating}

\usepackage{fancyhdr}
\graphicspath{{./eps/}}
\DeclareGraphicsExtensions{.eps,.eps.gz,.ps,.ps.gz}
\newcommand{\hdick}{\noalign{\hrule height1.4pt}}
\newcommand{\eV}  {\mathrm{eV}}

\newcommand{\MeV} {\mathrm{MeV}}
\newcommand{\GeV} {\mathrm{GeV}}
\newcommand{\TeV} {\mathrm{TeV}}
\newcommand{\fb}  {\mathrm{fb}}
\newcommand{\fbi} {\mathrm{fb}^{-1}}

\newcommand{\abi} {\mathrm{ab}^{-1}}

\newcommand{\cm}  {\mathrm{cm}}
\newcommand{\sek} {\mathrm{s}}
\newcommand{\msek}{\mathrm{ms}}
\newcommand{\cO}  {{\cal O}}
\newcommand{\cL } {{\cal L}}
\newcommand{\cP } {{\cal P}}

\def\susy    {{\sc Susy}}
\def\pythia  {{\sc Pythia}}
\def\suspect {{\tt SuSpect}}
\def\spheno  {{\tt SPheno}}

\def\tesla  {{\sc Tesla}}
\def\ilc    {{\sc ILC}}
\def\lhc    {{\sc LHC}}

\def\ee    {e^+e^-}
\def\ti    {\tilde}
\def\sa    {{\ti a}}

\def\stau  {{\ti\tau}}
\def\snu   {{\ti\nu}}
\def\sell  {{\ti\ell}}

\def\cx    {{\ti\chi}}
\def\ch    {{\ti\chi}}
\def\nt    {{\ti\chi^0}}
\def\sg    {{\ti g}}
\def\sG    {{\ti G}}

\def\smur  {{\ti\mu}_R}

\def\se    {{\ti e}}
\def\sel   {{\ti e}_L}
\def\ser   {{\ti e}_R}

\def\snu   {{\ti\nu}}

\def \Eslash {E \kern-.6em\slash }
\def \Mslash {M \kern-.5em\slash }

\newcommand{\nn}{\nonumber}
\newcommand{\eq}[1]{eq.~(\ref{#1})}
\newcommand{\fig}[1]{figure~\ref{#1}}
\newcommand{\tab}[1]{table~\ref{#1}}

\begin{document}

\title{  
  {\rm \large \hfill DESY 06-075 \\[-2.ex] \hfill May 2006} \\[1.ex]
  Detecting metastable staus and gravitinos at the ILC }
\author{ H.--U.~Martyn }   
\affiliation{I. Physikalisches Institut, RWTH Aachen, Aachen, Germany} 
\affiliation{Deutsches Elektronen-Synchrotron DESY, Hamburg, Germany}


\begin{abstract}
A study of various \susy\ scenarios is presented 
in which the lightest supersymmetric particle 
is the gravitino $\sG$ and the next-to-lightest supersymmetric particle 
is a scalar tau $\stau$ with lifetimes ranging from seconds to years. 
Gravitinos are interesting dark matter candidates which can be produced 
in decays of heavier sparticles at the International Linear Collider (\ilc), 
but remain undetected in direct searches of astrophysical experiments.
We investigate the detection and measurement of metastable staus, which may be
copiously produced at the \ilc\ either directly or via cascade decays.  
A proper choice of the experimental conditions will allow one to stop large 
samples of $\stau's$ in the
calorimeters of the \ilc\ detector and to study the subsequent decays
$\stau\to\tau\sG$. 
Detailed simulations show that the properties of the stau
and the gravitino, such as $\stau$ mass and lifetime and 
$\sG$ mass, can be accurately determined at a future \ilc\
and may provide direct access to the gravitational coupling, 
respectively Planck scale.
\end{abstract}

\maketitle

\section{Introduction}
  \label{intro} 

Supersymmetry (\susy) provides a very attractive scenario to account for the
amount of dark matter in the universe.
If $R$-parity is conserved, the lightest supersymmetric particle (LSP) is
stable and thus an ideal dark matter candidate.
Most frequently discussed is a neutralino LSP. 
Another interesting candidate is the spin $3/2$ 
gravitino $\sG$, cf. ref.~\cite{pagels},
which is part of the supersymmetric particle spectrum
in extensions to supergravity, or local \susy.
The mass of the gravitino is set by the \susy\ breaking scale $F$     
    \begin{eqnarray} m_{3/2} \ = \ 
      m_{\sG} & = & \frac{F}{\sqrt{3}\, M_P} \; , 
      \label{eq:gravitinomass}
    \end{eqnarray}   
where
$M_P = (8\pi\, G_N)^{-1/2} \simeq 2.4 \cdot 10^{18}~\GeV$ is the reduced 
Planck scale    
and $G_N$ is Newton's constant.
The masses of the other superparticles depend on the supersymmetry breaking 
mechanism and on the parameters of a specific model.
In general $m_{3/2}$ 
is a free parameter in the theory and may
extend over a wide range of $\cO(\eV-\TeV)$.

If the gravitino is indeed the LSP, it may be produced in cascade decays of
heavy \susy\ particles. 
To be specific, the next-to-lightest supersymmetric particle (NLSP) is
assumed to be a charged slepton, the scalar tau $\stau$.
The dominant decay will then be $\stau\to\tau\sG$ and since the coupling 
is gravitational, the lifetime may be very long, ranging from
seconds to years.
The decay-width $\Gamma_\stau$ 
of the metastable $\stau$ NLSP is given by
\begin{eqnarray}
  \Gamma_{\stau\to \tau\sG}  & = & \frac{1}{48\pi M_P^2}
  \frac{m_{\stau}^5}{m_{\sG}^2}
  \left[1 -\frac{m_{\tilde{G}}^2}{m_{\stau}^2} \right]^4 \ .
  \label{eq:decaywidth}
\end{eqnarray}
This formula will be the basis of the present study.
The $\stau$ decay rate, respectively lifetime $t_\stau=\Gamma_\stau^{-1}$,
depends only on the masses $m_\stau$ and $m_\sG$ as well as on the Planck 
scale $M_P$ -- no further supersymmetry parameters are required.
Conversely, the `supergravity Planck scale' 
can be expressed in terms of observable quantities
\begin{eqnarray}  
  M_P & = &
  \sqrt{ \frac{t_{\stau}\,m_\stau}{48\pi} }
  \frac{m_{\stau}^2}{m_{\sG}}
  \left[1 -\frac{m_{\tilde{G}}^2}{m_{\stau}^2} \right]^2 \ . 
  \label{eq:planckscale}
\end{eqnarray}
A comparison of such a measurement at a particle collider experiment 
with the value determined in macroscopic gravitational experiments,
$M_P = 2.436(2)\cdot 10^{18}\,\GeV$~\cite{pdg}, 
would be a unique test of supergravity~\cite{buchmueller}.
If the gravitational coupling of the $\stau$ decay is assumed, the direct
measurement of the gravitino mass may be replaced by a more accurate 
determination from the decay-width.
Rewriting \eq{eq:decaywidth},
the $\stau$ lifetime as a function of the stau and gravitino masses
can be expressed to a good approximation as
 \begin{eqnarray}
   t_{\stau} & \simeq & 3.6\cdot 10^7\,\sek
   \left[\frac{100\,\GeV}{ m_{\stau} - m_{\sG}}\right]^4
   \left[\frac{m_{\sG}}{100\,\GeV}\right] \ . 
   \label{eq:lifetime}
 \end{eqnarray}

The cosmological production of gravitino dark matter proceeds 
essentially via two mechanisms:
thermal production in 2-body QCD interactions,
{\it e.g.} $g g \to \sg\sG$ or $q \sg \to q \sG$,
in the reheating phase following inflation~\cite{bolz};
and/or late decays of the next-to-lightest sparticle, 
usually a slepton or neutralino,
which freezes out as usual and decays only after
the big bang nucleosynthesis phase, 
the prediction of which should not be destroyed
by the hadronic decays~\cite{frt}.
Comparing the calculated gravitino relic density with the observed 
dark matter density puts constraints on the NLSP and gravitino masses
and on the parameters for supersymmetric models.
Collider experiments offer a unique possibility to detect metastable staus
and to study the properties of gravitinos, 
which otherwise cannot be observed directly in astrophysical experiments.
Some phenomenological investigations to detect long-lived staus
and to measure their lifetime
at the \lhc\ and \ilc\ are discussed in \cite{hamaguchi, fs,gdm}.

In the present study the excellent potential of future \ilc\ experiments 
and detectors
will be explored to observe and detect metastable $\stau's$, 
measure precisely their mass and lifetime, 
and to determine in an independent way the gravitino mass.
In section~2 the investigated physics scenarios will be presented. 
Details on experimental aspects, detector requirements and event simulation
will be given in sections 3 and 4. 
The experimental analyses and case studies are described in section~5.
The results are discussed in section~6, followed by conclusions.

\section{Physics scenarios}
  \label{scenarios}

Several \susy\ benchmark models which have been recently discussed in the 
literature are chosen for the case studies.
They represent different supersymmetry breaking mechanisms and a variety
of stau and gravitino masses leading to different experimental conditions.
In general, the models are consistent with cosmological constraints,
as claimed by the authors.
Here they should just serve as valid examples of `typical' spectra.
The gravitino masses are somewhat arbitrary. 
However, most important, the results can be easily transformed to other 
scenarios with different stau and gravitino masses.
The masses of the sleptons, light neutralinos and charginos, relevant for
the benchmark models under study, are compiled in \tab{tab:gdmspectra}.

{\it Gauge mediated symmetry breaking} (GMSB) 
usually occurs at rather low scales and a light gravitino is naturally 
the LSP~\cite{dine}.
Typical masses are of order eV to keV which may, however, 
be extended in the GeV range~\cite{moroi}.
The GMSB scenario SPS~7 from the Snowmass points~\cite{sps} is
chosen as reference model. 
It is described by the conventional parameters
$\Lambda=40\,\TeV$, $M_m=80\,\TeV$, $N_m=3$, $\tan\beta=15$ and
sign\,$\mu = +$,
where $\Lambda$ is
the universal soft \susy\ breaking scale, 
$M_m$ and $N_m$ are the messenger scale and index,
$\tan\beta$ is the ratio of the vacuum expectation values of the two 
Higgs fields, and
sign\,$\mu$ is the sign of the higgsino mass parameter $\mu$.
The gravitino mass is set arbitrarily to $m_{\sG}=0.1\,\GeV$.

In {\it supergravity mediated symmetry breaking} (SUGRA) the gravitino mass
$m_{3/2}$ is a free parameter and of the same order as the other sparticle
masses~\cite{nilles}.
If the NLSP is required to be the charged $\stau$, 
then the common scalar mass $m_0$ in minimal versions
has to be small and much lower than the common gaugino mass $M_{1/2}$.
In the scenario FS~600 of ref.~\cite{fs}
the mSUGRA parameters are
$m_0=0\,\GeV$, $M_{1/2}=600\,\GeV$,
$A_0=0\,\GeV$, $\tan\beta=10$ and sign\,$\mu = +$,
where $A_0$ is the common trilinear coupling.
The gravitino mass is chosen to be $m_{\sG}=50\,\GeV$.
The authors of the GDM series~\cite{gdm} require a tighter mSUGRA definition 
with unified scalar and gravitino masses $m_{3/2} = m_0$.
The parameters of GDM~$\epsilon$ are
$m_0=m_{3/2}=20\,\GeV$, $M_{1/2}=440\,\GeV$,
$A_0=25\,\GeV$, $\tan\beta=15$ and sign\,$\mu = +$.
The model GDM~$\zeta$ is given by
$m_0=m_{3/2}=100\,\GeV$, $M_{1/2}=1000\,\GeV$,
$A_0=127\,\GeV$, $\tan\beta=21.5$ and sign\,$\mu = +$.
The spectrum of GDM~$\eta$ differs from the previous one
essentially only by a lighter gravitino, thus
$m_0=m_{3/2}=20\,\GeV$, $M_{1/2}=1000\,\GeV$,
$A_0=25\,\GeV$, $\tan\beta=23.7$ and sign\,$\mu = +$.

In {\it gaugino mediated symmetry breaking} ($\ch$MSB) the gravitino may be
the LSP with a mass comparable to the other sparticles~\cite{kaplan}.
In the scenarios proposed by~\cite{bks}
a gravitino LSP can be accompanied by a stau, sneutrino or 
neutralino NLSP, where the latter options are essentially unobservable.
The spectra of the $\stau$ NLSP models are determined by the parameters
$M_{1/2}=500\,\GeV$, $\tan\beta=10$, sign\,$\mu = +$ 
and the soft Higgs masses,
chosen as $m_{\tilde h_1}^2=m_{\tilde h_2}^2=0\,\TeV^2$ for point $\ch$MSB~P1,
and $m_{\tilde h_1}^2=0\,\TeV^2$, $m_{\tilde h_2}^2=0.5\,\TeV^2$ 
for point $\ch$MSB~P3.
An interesting feature of this particular gaugino mediation scenario
is that a naive dimensional analysis provides a lower
bound on the gravitino mass of $m_{3/2}\gtrsim 10\,\GeV$~\cite{bhk}.
In the present case studies  
values of $m_\sG=50\,\GeV$ and $m_\sG=25\,\GeV$ are chosen for P1 and P3,
respectively.

The sparticle spectra of \tab{tab:gdmspectra}
are calculated from the original set of parameters using the programs
 \spheno~\cite{spheno} for models 1, 2 and 
   \suspect~\cite{suspect} for models  3 -- 7.

\renewcommand{\arraystretch}{1.1}
\begin{table}[tb] \centering 
  \caption{Spectra of sleptons, light neutralinos and charginos
    [masses in GeV] in \susy\ scenarios with a gravitino LSP, 
    masses of the first and second generation are identical;
    the gravitino masses [GeV] used in the case studies are given in the
    last line}
  \label{tab:gdmspectra}
  \vspace{3mm}
  \begin{tabular}{ l c c c c c c c}
    \hdick  
    \phantom{xxxxxxx}& \phantom{xxxxxxxxxx}& \phantom{xxxxxxxxxx}
                     & \phantom{xxxxxxxxxx}& \phantom{xxxxxxxxxx}
                     & \phantom{xxxxxxxxxx}& \phantom{xxxxxxxxxx}
                     & \phantom{xxxxxxxxxx} 
    \\ [-3.ex]
    & 1 & 2 & 3 & 4 & 5 & 6 & 7 \\
    $\sell$, $\cx$, $\sG$  
                    & SPS 7 & FS 600 & GDM $\epsilon$ & GDM $\zeta$ 
                                     & GDM $\eta$
                    & $\ch$MSB P1    & $\ch$MSB P3                 
    \\[.5ex] \hdick
    $\stau_1$   & 123.4 & 219.3 & 157.6 & 340.2 & 322.1 & 185.2 & 102.5 \\
    $\stau_2$   & 264.9 & 406.5 & 307.2 & 659.2 & 652.2 & 341.5 & 356.9 \\
    $\snu_\tau$ & 249.6 & 396.4 & 290.9 & 649.5 & 641.5 & 327.7 & 346.9 \\
    \hline
    $\ser  $    & 130.9 & 227.2 & 175.1 & 381.4 & 368.5 & 192.9 & 109.5 \\ 
    $\sel  $    & 262.8 & 405.6 & 303.0 & 662.7 & 655.3 & 340.1 & 357.4 \\ 
    $\snu_e$    & 250.1 & 397.6 & 292.8 & 658.1 & 650.7 & 328.0 & 247.4 \\ 
    \hline
    $\nt_1 $    & 163.7 & 243.0 & 179.4 & 426.5 & 426.5 & 203.6 & 189.2 \\  
    $\nt_2 $    & 277.9 & 469.6 & 338.2 & 801.9 & 801.5 & 385.5 & 263.8 \\  
    $\ch_1^\pm$ & 275.5 & 469.9 & 338.0 & 801.9 & 801.4 & 388.2 & 251.5 \\  
    \hline
    $\sG  $     &   0.1 &  50   &  20   & 100   &  20   &  50   &  25   \\
    \hline
  \end{tabular}    
\end{table}

\section{Stau detection \& measurement principles}

The programme to determine the properties of the metastable stau 
and to measure, independently, the mass of the gravitino
consists of several parts: \\[.5ex] \indent
--
    identify $\stau$ by the characteristic heavy ionisation 
    $-dE/dx \propto 1/\beta^2$  in the TPC; \\[.5ex] \indent
--
    determine $\stau$ mass from two-body kinematics of $\ee\to\stau_1\stau_1$ 
    production; \\[.5ex] \indent
--
    follow low momentum
    $\stau$ candidate until it stops inside detector,
    record $\stau$ location and time stamp $t_{0}$; \\[.5ex] \indent
--
    trigger decay $\stau\to\tau\sG$ at $t_{trig}$
    uncorrelated to beam collision; \\[.5ex] \indent
--
    associate vertex to previously recorded stopping point, 
    get $\stau$ lifetime $t_{\stau} = t_{trig}-t_{0}$; \\[.5ex] \indent
--
    measure $\tau$ recoil spectra in calorimeter, get
    gravitino mass $m_{\sG}$.   \\[.5ex] 
The requirements on the detector performance and operation will be discussed
in the following.

\begin{figure}[tb]
\setlength{\unitlength}{1.37mm} 
\begin{picture}(150,83)(50,15)
  \put(-10,0){
    \put(78,20){\epsfig{file=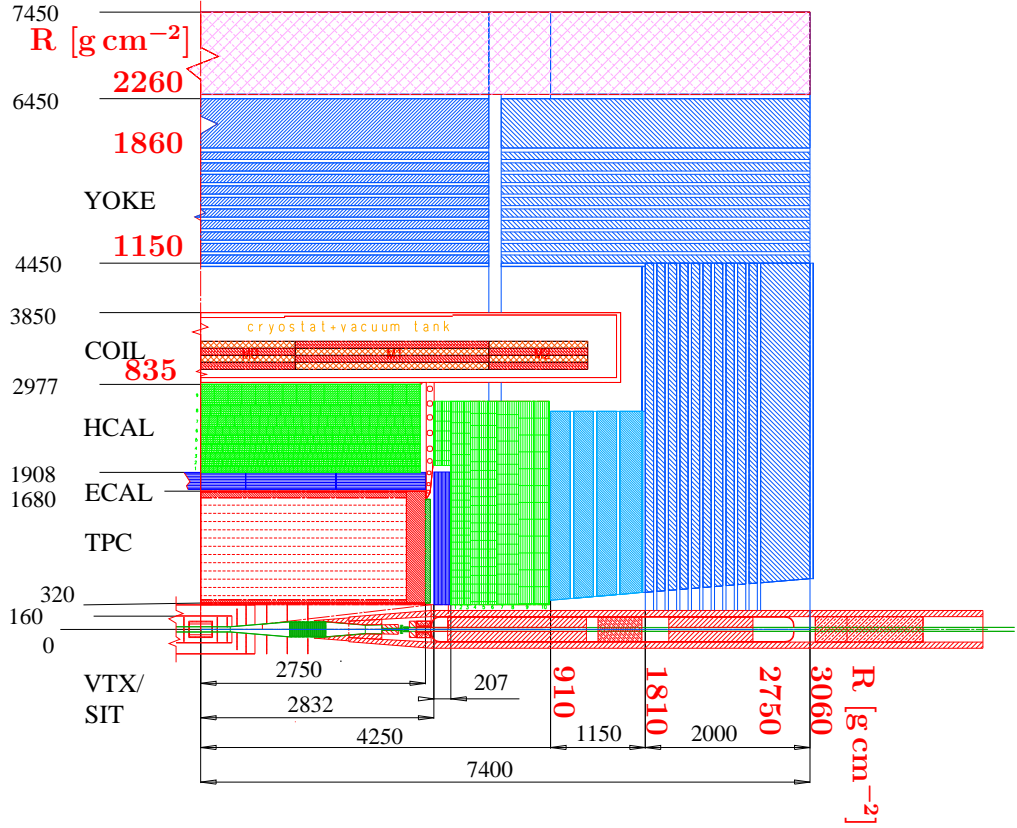,width=134mm}} 
    \color{red} \boldmath\bf\large 
    \put(87,-.5){
      \put(0,60){~~835} 
      \put(0,72){ 1150}  
      \put(0,82){ 1860}  
      \put(0,88){ 2260}  
      \put(-8,92){ R~[g\,cm$^{-2}$]} }
    \begin{rotate}{-90} 
      \put(-33,0){
        \put(0,131){  910}        
        \put(0,140){ 1810}        
        \put(0,151){ 2750}        
        \put(0,156){ 3060}        
        \put(0,160){ R~[g\,cm$^{-2}$]} }
    \end{rotate}  
  }
\end{picture}
\caption{View of one quadrant of the \tesla\ detector~\cite{tdr},
  length units in mm;
  the amount of material integrated perpendicular and along the 
  beam direction is indicated by slightly larger (red) symbols as
  $R~[{\rm g}\,\cm^{-2}]$}
\label{fig:detector}
\end{figure}

The detector is taken from the {\sc Tesla tdr}~\cite{tdr}; a cross section 
through one quadrant is displayed in \fig{fig:detector}.
This concept serves as a baseline for 
a future LDC detector~\cite{ldc}. 
In particular the calorimeter layout and depth 
may be taken as representative for various
detector designs discussed in the ILC community.
The main characteristics of the detector, relevant to the present study, are:
\begin{itemize}
  \item[--] A TPC with excellent tracking performance,  {\it i.e.}
    momentum resolution $\delta(1/p_t) < 2\cdot10^{-4} \,\GeV^{-1}$
    and dE/dx resolution $<5\,\%$. 
  \item[--] A highly segmented hadronic calorimeter (HCAL),
    40 samplings with $3 \times 3\,\cm^2$ lateral readout cells,  
    leading to energy resolutions of 
    $\delta E_{had} = 0.5\sqrt{E / \GeV}\,\oplus\, 0.04\,E$ for hadrons and
    $\delta E_{em} = 0.2 \sqrt{E / \GeV}\,\oplus\, 0.02\,E$ 
    for electromagnetic showers.
  \item[--] An instrumented iron flux return yoke equipped with several 
    layers for muon detection as well as for calorimetric measurements of
    hadron showers.
\end{itemize}

The amount of absorber material integrated along the radial and 
longitudinal directions,  $R\,[{\rm g/cm}^2]$,
can be read off the detector \fig{fig:detector}.
The range of a heavy, non-strongly interacting charged particle of 
mass $m$, is conveniently expressed
by $R/m\,[{\rm g}\,\cm^{-2}\GeV^{-1}]$ and can be parametrised 
as a function of the scaled momentum $p/m=\beta\gamma$
\begin{eqnarray}
   \log_{10} (R/m) & =& c_1 + c_2\log_{10}(p/m) \, ,
\end{eqnarray}
with $c_1= 2.087$ and $c_2=3.227$ for steel~\cite{rossi}.
Values of $\beta\gamma$ below which a heavy $\stau$ can be stopped in the 
hadron calorimeter or instrumented iron yoke depend on the 
location and angle of incidence. 
Typical ranges are listed in \tab{tab:bgranges}.
\begin{table}[htb]
  \centering
  \caption{Acceptance ranges of maximal $\beta\gamma$ at which heavy $\stau's$ 
    can be absorbed in the HCAL or instrumented iron yoke of the ILC detector}
  \label{tab:bgranges}
  \renewcommand{\arraystretch}{1.1}
  \begin{tabular}{c c c} 
    $m_\stau  $  & ~~ $\beta\gamma$ (HCAL) ~~ & ~~ $\beta\gamma$ (Yoke) ~~
    \\[.5ex] \hdick
    $125\;\GeV$  & $0.41 - 0.46$ & $0.52 - 0.59$ \\
    $250\;\GeV$  & $0.33 - 0.37$ & $0.42 - 0.48$ \\
    $375\;\GeV$  & $0.29 - 0.33$ & $0.37 - 0.41$ \\
  \end{tabular}    
\end{table}

\section{Event generation}
  \label{eventgeneration}

Events are generated with the program \pythia~6.3~\cite{pythia}
which includes initial and final state QED radiation as well as 
beamstrahlung~\cite{circe}.
The detector simulation is based on the detector proposed
in the {\sc Tesla tdr}~\cite{tdr}
and implemented in the Monte Carlo program {\sc Simdet}~4.02~\cite{simdet}.

The analysis of \susy\ scenarios with a gravitino LSP is quite different
from conventional supersymmetric models. The metastable $\stau$ NLSP 
will be detected before it leaves or will be absorbed in the detector.
Thus, there is a very clean signature without missing particles 
(except neutrinos from decays): the observed particle momenta are balanced
but their moduli don't sum up to the cms energy
\begin{eqnarray}
   \left |\; \sum_i \vec{p_i}\; \right  | \simeq 0 
   \quad &{\rm and } & \quad
  \sum_i p_i < \sqrt{s}\; .  \nn
\end{eqnarray}
These features are very distinct from Standard Model background which can
be efficiently rejected. 
They also allow the sparticle production and decay chains to be reconstructed
from the complete event kinematics.
Each \susy\ event contains two $\stau's$ which can be readily identified as 
highly ionising tracks and their passage through the detector can be
accurately followed. The location of stopping $\stau's$ may be determined
within a volume of a few $\cm^3$.

The decay $\stau\to\tau\sG$ is triggered by an isolated, high energy
hadronic or electromagnetic cluster in the HCAL 
(threshold $E_{h,em}>10\,\GeV$),
by a hadronic shower in the yoke ($E_{h}>10\,\GeV$)
or by an energetic $\mu$ originating in the HCAL or yoke ($E_\mu>10\,\GeV$), 
occurring at any time not correlated to beam collisions.
The main background is expected to come from cosmic rays 
and can be rejected by excluding decay vertices in the outermost detector 
layers or signals being initiated by muons from outside.
Further excellent background discrimination is provided by requiring the 
decay vertex to coincide with a previously recorded $\stau$ stopping point.
The $\tau$ decay modes contributing to the analysis are
the leptonic 3-body decays 
$\tau\to\mu\nu_\mu\nu_\tau$ (17.4\%) and
$\tau\to e\nu_e\nu_\tau$ (17.8\%), 
and the hadronic decays 
$\tau\to\pi\nu_\tau$ (11.1\%),  
$\tau \to\rho\nu_\tau 
     \to\pi^\pm\pi^0\nu_\tau$ (25.4\%) and
$\tau \to 3\pi\nu_\tau 
     \to \pi^\pm \pi^+\pi^-\nu_\tau + 
  \pi^\pm\pi^0\pi^0 \nu_\tau$ (19.4\%).

The $\tau$ recoil energy gives access to the gravitino mass
\begin{eqnarray}
   E_\tau & = & \frac{m_\stau}{2}\, 
   \left ( 1 - \frac{m_\sG^2}{m_\stau^2} \right )
   \qquad\qquad {\rm for ~~~ } m_\tau^2 \ll m_\stau^2 \ .
   \label{eq:etau}
\end{eqnarray}
The sensitivity to low gravitino masses decreases rapidly, e.g. 
for $m_\sG = 0.1\,m_\stau$ and $m_\stau$ precisely known
the energy $E_\tau$,
which is also the maximum energy of the observable
$\tau$ decay products,
has to be measured with a precision well below one percent.

\section{Experimental analyses -- case studies}
\label{analyses}

A big advantage of the \ilc\ is that the centre of mass energy can be adjusted 
in order to control the sparticle production and the
number of staus trapped in the detector.
A range-out in the calorimeter is preferred for measuring the kinematics 
of $\stau$ decays and thus the gravitino mass, while all staus
stopped somewhere in the detector can be used for determining the lifetime.
Low momentum $\stau's$ with a suitable $\beta\gamma$ factor
can be produced either directly or via cascade decays from light sleptons,  
e.g. $\ser\to e\tau\stau_1$,
or neutralinos, $\nt_1\to\tau\stau_1$.
However, all these processes 
--- $\stau_1\stau_1$, $\ser\ser$, $\smur\smur$ and $\nt_1\nt_1$ ---
rise only slowly above kinematic threshold with cross sections 
$\sigma\propto\beta^3$, thus providing relatively low rates.
Obviously, choosing $\stau_1\stau_1$ pair production as the only 
source is not optimal.
A more efficient reaction, if kinematically accessible, is the associated
selectron production $\ee\to\ser\sel\to e\tau\stau_1\, e\nt_1$, 
which increases much faster as $\sigma\propto\beta$ near 
threshold~\cite{freitas}.
Note, such a threshold behaviour also applies to 
$e^-e^-\to \ser\ser$ production~\cite{peskin}.

For the present studies the collider energies are not always optimised.
Instead, the often proposed `standard' energies of 
500\,GeV (models 2, 3, 7) and 800\,GeV (models 4, 5) are assumed,
as well as some special settings around 420\,GeV (models 1, 6).
The integrated luminosities are moderate and chosen such as to enable
a reasonable measurement of the $\stau$ decay spectra.
The results obtained may be easily scaled to apply to different 
running conditions, event statistics and detector efficiencies.

The experimental analysis and techniques will be discussed 
representatively in some detail for the GDM~$\epsilon$ scenario. 
For the other benchmark models the main features and results will be 
presented in a more compact form.
All results are summarised in \tab{tab:results}.

\subsection{mSUGRA scenario {\boldmath GDM~$\epsilon$} }

The assumptions for the case study of the GDM~$\epsilon$ scenario consist
of the observables 
$m_\stau=157.6\,\GeV$, $t_\stau=2.6\cdot10^6\,\sek$, $m_\sG=20\,\GeV$
and the experimental conditions of 
$\sqrt{s}=500\,\GeV$ centre of mass energy, 
$\cL= 100\,\fbi$ integrated luminosity,
and $\sigma_{SUSY} = 300 \,\fb$ as inclusive $\stau$ production cross section.

The scaled momentum distribution $\beta\gamma$ of $\stau's$ produced
in various reactions is shown in \fig{fig:gdme_all}\,a. 
The majority of particles, 
coming from diagonal slepton and neutralino pairs and
peaking around $\beta\gamma\simeq 1$, will leave the detector. 
One observes, however, a second peak at low $\beta\gamma\lesssim 0.5$
which stem from cascade decays of $\ser\sel$ production 
and which may be stopped inside the detector, 
see $\beta\gamma$ ranges in \tab{tab:bgranges}. 
One expects $N_\stau^{\rm hcal} = 4100$
and $N_\stau^{\rm yoke} = 1850$ $\stau's$ trapped in the calorimeter and
yoke, respectively.

\begin{figure}[htb]
  \setlength{\unitlength}{1mm} 
  \begin{picture}(150,140)(0,0) 
    \put(75,70){
      \put(0,0){\epsfig{file=mepsilon_fitspectra.eps,angle=90,width=80mm}} 
      \put(12,64){\epsfig{figure=box.eps,height=5mm,width=50mm}}
      \put(68,49){ \large\bf b) }      
      \put(50,58){ $m_\stau=158\,\GeV $}      
      \put(50,55){ $m_\sG=~20\,\GeV $}      
      \put(45,0){\epsfig{figure=box.eps,height=6mm,width=50mm}}
      \put(45,0){\large $\tau$ jet $E_{jet}~[\GeV] $}  }    
    \put(-15,70){
      \put(-8,-5){\epsfig{file=mepsilon_betagamma.eps,angle=90,width=91mm}}
      \put(68,49){ \large\bf a) }      
      \put(52,58){{\color{red}\large\bf\boldmath GDM $\epsilon$}}
      \put(20,-3){\large
        \put(-5,53){\epsfig{figure=box.eps,height=16mm,width=20mm}}
        \put(0,55){\color{magenta} $\stau_1\stau_1$}
        \put(0,50){\color{cyan} $\ser\se_{R,L}$}
        \put(0,45){\color{yellow} $\smur\smur$}
        \put(0,40){\color{green} $\nt_1\nt_1$} }
      \put(50,-5){\epsfig{figure=box.eps,height=6mm,width=40mm}}
      \put(50,0){\large $p_\stau/m_\stau = \beta\gamma$}  
      \put(15,58){\color{black}\large \bf\boldmath $\sqrt{s}=500\,\GeV$} }
    \put(75,0){
      \put(0,0){\epsfig{file=t_vsm.eps,angle =90,width=80mm}}      
      \put(68,49){ \large\bf d) } }
    \put(-15,0){
      \put(0,0){\epsfig{file=mepsilon_lifetime.eps,angle =90,width=80mm}}    
      \put(68,49){ \large\bf c) } }
  \end{picture}
  \caption{ a) Scaled momentum spectrum $p/m = \beta\gamma$ of $\stau$ 
    production
    with contributions from $\stau_1\stau_1$ (magenta), $\ser\se_{R,L}$ (cyan),
    $\smur\smur$ (yellow) and $\nt_1\nt_1$ (green);
    b) $\tau$ recoil spectrum of the decay $\stau_1\to\tau\sG$ with
    $E_{jet}$ of $\tau\to\rho\nu_\tau + 3\pi\nu_\tau$ compared 
    with simulations assuming $m_\sG=20\,\GeV$ (red histogram), 
    $10\,\GeV$ (blue) and $30\,\GeV$ (magenta);
    c) $\stau$ lifetime distribution; 
    d) $\stau$ lifetime versus $\sG$ mass for various \susy\
    scenarios representing different $\stau$ masses.
    Statistics correspond to
    GDM $\epsilon$ scenario, $\cL = 100\,\fbi$ at $\sqrt{s} = 500\,\GeV$}
  \label{fig:gdme_all}
\end{figure}

The {\it stau mass} measurement is based on the kinematics of prolific
pair production $e^+e^-\to\stau_1\stau_1$, 
see magenta curve in \fig{fig:gdme_all}~a, 
to be identified as a pair of collinear, non-interacting particles with 
momenta  $p_\stau<\sqrt{s}/2 = E_\stau$, 
{\it i.e.} well below the beam energy. 
Determining the mean value of the $\stau$ momentum with an accuracy of
$ \langle p_\stau \rangle = 192.4 \pm 0.2\;\GeV$ 
leads to a precise $\stau$ mass of 
\begin{eqnarray}
  m_\stau = 157.6\pm 0.2\;\GeV\ . \nn
\end{eqnarray}

The {\it stau lifetime} measurement is based on the decays of $\stau's$
which have been stopped in the detector. 
Requiring an isolated energetic cluster above a threshold of $10\,\GeV$
or a muon above $10\,\GeV$ originating somewhere inside the sensitive
fiducial volume of the calorimeter or yoke,
results in the decay time distribution shown in \fig{fig:gdme_all}~c.
A fit to the spectrum gives a $\stau$ lifetime of 
\begin{eqnarray}
  t_\stau &=& (2.6\pm0.05)\cdot10^6\,\sek \ , \nn
\end{eqnarray}
corresponding to roughly one month.
The relation between stau lifetime and gravitino mass
as a function of stau mass 
is graphically presented in \fig{fig:gdme_all}~d.
For the present condition the lifetime increases as $t_\stau \sim m^2_\sG$
and rises significantly faster for $m_\sG \gtrsim 0.25\,m_\stau$, 
reaching $10^8\,\sek$ for a $75\,\GeV$ gravitino.
Comparing with other physics scenarios
one also observes that the lifetime can get very long for light $\stau's$
and small mass differences
and, for a fixed gravitino mass, becomes shorter the heavier the stau.

A {\it direct gravitino mass} measurement can be performed by exploiting 
the $\tau$ recoil of the decay $\stau\to\tau\sG$,
see \eq{eq:etau} and discussion in sect.~\ref{eventgeneration}.
The upper endpoints of the energy spectra 
which coincide with the primary $\tau$ energy $E_\tau=77.5\,\GeV$,
are directly related to the masses involved.
The leptonic 3-body decay $\tau\to e\nu\nu$ is not very useful due to
the soft spectrum peaking at low values.
The hadronic 2-body decay $\tau\to\pi\nu$ produces a flat spectrum, 
where the upper edge is, however, strongly diluted due to resolution effects 
and limited statistics.
Well defined upper edges are provided by the hadronic decays to heavier final 
states $\tau\to\rho\nu$ and $\tau\to\pi\pi\pi\nu$.
The energy distribution of both decay modes, defined as `$\tau$ jets',
is shown in \fig{fig:gdme_all}~b. 
In order to illustrate the sensitivity to the gravitino mass
simulations assuming the nominal value of $m_\sG=20\,\GeV$ and values shifted
by $\pm 10\,\GeV$ are shown as well. 
A fit to the 'jet' energy spectrum, 
using either an analytical formula for the endpoint or 
a complete simulation, yields a gravitino mass
\begin{eqnarray}  
  m_\sG&=&20\pm 4\,\GeV \ .
\end{eqnarray}

Taking all results together one can test the gravitational coupling of 
the stau to the gravitino and access the Planck scale, 
respectively Newton's constant.
Inserting the expected values and accuracies
on $m_\stau$, $t_\stau$ and $m_\sG$ in \eq{eq:planckscale}
one finds for the supergravity Planck scale
\begin{eqnarray}  
  M_{P} & = & (2.4 \pm 0.5)\cdot 10^{18}\;\GeV \ , \nn
\end{eqnarray}
where the error is dominated
by the precision on the gravitino mass measurement.

The {\it gravitino mass} can be deduced more precisely
from the $\stau$ mass and lifetime,
if the gravitational coupling is assumed and the macroscopic value of
$M_P$ is used in the decay-width of \eq{eq:decaywidth}.
The resulting gravitino mass is $m_\sG=20\pm 0.2\,\GeV$, 
where the error is dominated by the lifetime measurement.

It is a unique feature of gravitino LSP scenarios that the Planck scale can 
be directly measured in microscopic particle experiments by studying
the properties of the NLSP and its decay.
A further interesting test to reveal the nature of the gravitino
as the supersymmetric partner of the graviton
would be to determine the spin. 
This is in principle possible by studying correlations in the radiative decay
$\stau\to\tau\gamma\sG$~\cite{buchmueller}, see also sect.~\ref{discussion}.
Experimentally this is quite challenging, because
it requires to distinguish single photons from $\pi^0$ decays,
and in addition the expected rates are lower by two orders of magnitude.

\subsection{mSUGRA scenario {\boldmath GDM $\zeta$} }

Due to the heavy sleptons in model GDM~$\zeta$ a higher
centre of mass energy of $\sqrt{s}=800\,\GeV$ 
and an integrated luminosity of $\cL=1\,\abi$ is chosen.
The $\beta\gamma$ distribution of the produced $\stau's$ 
(inclusive cross section $\sigma_{SUSY}=5\,\fb$)
is shown in \fig{fig:gdmzeta_spectra}~a.
It exhibits the typical behaviour of slepton production close to threshold:
a pronounced peak at $\beta\gamma\simeq 0.6$ from $\stau_1\stau_1$ pairs
and another enhancement around 0.3 from $\ser$ and $\smur$ cascade decays,
resulting in 1350 (850) trapped $\stau's$ in the hadron calorimeter (yoke).
The analyses of the momentum spectrum yields a $\stau$ mass of
$m_{\stau_1}=340.2\pm 0.2\,\GeV$.
From the decay time distribution one gets the lifetime
$t_\stau=(1.8\pm 0.06)\cdot 10^6\,\sek$.
The $\tau$ jet energy spectrum of the $\stau$ decay, 
shown in \fig{fig:gdmzeta_spectra}~b, extends over a wide range.
The edge of the endpoint energy, 
being less pronounced than in the previous example,
can be used to get a direct gravitino mass measurement 
of $m_{\sG}=100\pm 10\,\GeV$.
This value may be compared with the
gravitino mass $m_{\sG}=100\pm 2\,\GeV$ calculated from the stau mass and
lifetime.

{\it Note:} 
If higher energies were accessible at the \ilc, then data taking at 
$\sqrt{s}=900\,\GeV$, {\it i.e.} just above  $\nt_1\nt_1$ threshold,
increases the number of trapped $\stau's$ by a factor of 1.5 
in the GDM~$\zeta$ scenario.

\begin{figure}[htb]
  \setlength{\unitlength}{1mm} 
  \begin{picture}(150,70)(0,0) 
    \put(75,0){
      \put(0,0){\epsfig{file=mzeta_fitspectra.eps,angle=90,width=80mm}}
      \put(68,49){ \large\bf b) }      
      \put(12,64){\epsfig{figure=box.eps,height=5mm,width=50mm}}
      \put(50,58){ $m_\stau=340\,\GeV $}      
      \put(50,55){ $m_\sG=100\,\GeV $}      
      \put(45,0){\epsfig{figure=box.eps,height=6mm,width=50mm}}
      \put(45,0){\large $\tau$ jet $E_{jet}~[\GeV] $}  }    
    \put(-15,0){
      \put(0,0){\epsfig{file=mzeta_betagamma.eps,angle=90,width=80mm}}
      \put(68,49){ \large\bf a) }      
      \put(52,58){{\color{red}\large\bf\boldmath GDM $\zeta$}}
      \put(20,-3){\large
        \put(-7,53){\epsfig{figure=box.eps,height=16mm,width=20mm}}
        \put(0,55){\color{magenta} $\stau_1\stau_1$}
        \put(0,50){\color{cyan} $\ser\se_{R}$}
        \put(0,45){\color{yellow} $\smur\smur$}}
      \put(50,0){\epsfig{figure=box.eps,height=6mm,width=40mm}}
      \put(50,0){\large $p_\stau/m_\stau = \beta\gamma$}  
      \put(15,58){\color{black}\large \bf\boldmath $\sqrt{s}=800\,\GeV$} }
  \end{picture}
  \caption{ a) Scaled momentum spectrum 
    $p/m=\beta\gamma$ of $\stau$ production
    with contributions from $\stau_1\stau_1$ (magenta), $\ser\se_{R}$ (cyan)
    and $\smur\smur$ (yellow);
     b) $\tau$ recoil spectrum of the decay $\stau_1\to\tau\sG$ with
    $E_{jet}$ of $\tau\to\rho\nu_\tau + 3\pi\nu_\tau$ compared
    with simulations using $m_\sG=100\,\GeV$ (red histogram), 
    $80\,\GeV$ (blue) and $120\,\GeV$ (magenta).
    GDM $\zeta$ scenario, $\cL=1000\,\fbi$ at $\sqrt{s}=800\,\GeV$}
  \label{fig:gdmzeta_spectra}
\end{figure}

The {\it GDM~$\eta$ scenario} has an almost identical sparticle spectrum and
differs essentially only by the lower gravitino mass of $m_\sG=20\,\GeV$.
This causes a much shorter lifetime, but, more importantly,
the $\tau$ recoil energy spectrum is not really sensitive 
to the gravitino mass ($m_\sG/m_\stau = 0.06$)
and can only be used to set an upper limit.

\subsection{mSUGRA scenario FS 600}

The model FS~600 is investigated at a cms energy $\sqrt{s}=500\,\GeV$,
close to  threshold pair production of the moderately heavy sleptons
and neutralinos.
A larger integrated luminosity is needed in order to achieve similar 
accuracies as for the other mSUGRA benchmarks.
The results of the analysis are given in \tab{tab:results}.
It is worth noting that operating the \ilc\ at a slightly higher energy of
520~GeV leads to higher cross sections and a significant increase 
by a factor of 1.5 for the rate of trapped $\stau's$.

\subsection{Gauge mediated symmetry breaking scenario SPS 7}

The SPS~7 scenario has a relatively light sparticle spectrum and
is investigated assuming $\sqrt{s}=410\,\GeV$ and $\cL=100\,\fbi$,
with a large cross section of $\sigma_{SUSY}=420\,\fb$ for inclusive
$\stau$ production.
As can be seen in the $\beta\gamma$ distribution of \fig{fig:sps7_spectra}~a,
most $\stau's$ leave the detector.
There is, however, a large signal at $\beta\gamma\simeq0.4$ 
from $\ser\sel$ production just above threshold, 
which contributes to the sample of 
10000 (4900) trapped $\stau's$ in the calorimeter (yoke).
The analysis of the $\stau$ momentum spectrum yields 
$m_{\stau_1}=124.3\pm 0.1\,\GeV$ 
and from a fit to the decay time distribution one obtains
$t_\stau=209.3\pm \,2.4\,\sek$.
These values can be used to derive a very accurate gravitino mass of 
$m_\sG=100\pm 1\,\MeV$ {\it assuming} a gravitational coupling,
see \fig{fig:gdme_all}~d.
The $\tau$ recoil energy spectrum, shown in \fig{fig:sps7_spectra}~b, 
is not sensitive to such low gravitino masses and can only serve to give 
an upper limit on a direct measurement  
of $m_\sG < 9\,\GeV$ at 95\% confidence level.

\begin{figure}[htb]
  \setlength{\unitlength}{1mm} 
  \begin{picture}(150,70)(0,0) 
    \put(75,0){
      \put(0,0){\epsfig{file=sps7_fitspectra.eps,angle=90,width=80mm}} 
      \put(68,49){ \large\bf b) }      
      \put(12,64){\epsfig{figure=box.eps,height=5mm,width=50mm}}
      \put(50,58){ $m_\stau=124\,\GeV $}      
      \put(50,55){ $m_\sG=~0.1\,\GeV $}      
      \put(45,0){\epsfig{figure=box.eps,height=6mm,width=50mm}}
      \put(45,0){\large $\tau$ jet $E_{jet}~[\GeV] $}  }    
    \put(-15,0){
      \put(0,0){\epsfig{file=sps7_betagamma.eps,angle=90,width=80mm}}
      \put(68,49){ \large\bf a) }      
      \put(52,58){{\color{red}\large\bf\boldmath SPS 7}}
      \put(20,-3){\large
        \put(-7,53){\epsfig{figure=box.eps,height=16mm,width=20mm}}
        \put(0,55){\color{magenta} $\stau_1\stau_1$}
        \put(0,50){\color{cyan} $\ser\se_{R,L}$}
        \put(0,45){\color{yellow} $\smur\smur$}
        \put(0,40){\color{green} $\nt_1\nt_1$} }
      \put(50,0){\epsfig{figure=box.eps,height=6mm,width=40mm}}
      \put(50,0){\large $p_\stau/m_\stau = \beta\gamma$}  
      \put(15,58){\color{black}\large \bf\boldmath $\sqrt{s}=410\,\GeV$} }
  \end{picture}
  \caption{ a) Scaled momentum spectrum 
    $p/m = \beta\gamma$ of $\stau$ production
    with contributions from $\stau_1\stau_1$ (magenta), $\ser\se_{R,L}$ (cyan),
    $\smur\smur$ (yellow) and $\nt_1\nt_1$ (green);
     b) $\tau$ recoil spectrum of the decay $\stau_1\to\tau\sG$ with
    $E_{jet}$ of $\tau\to\rho\nu_\tau + 3\pi\nu_\tau$ compared  
    with simulations using $m_\sG=0\,\GeV$ (red histogram) and 
    $10\,\GeV$ (blue).
    SPS 7 scenario, $\cL=100\,\fbi$ at $\sqrt{s}=410\,\GeV$}
  \label{fig:sps7_spectra}
\end{figure}

In the present case one does not profit from the calorimetric information 
to asses the gravitino mass directly 
and relies entirely on the stau properties.
One may afford to run at higher energy,
thereby shifting the $\beta\gamma$ spectrum of \fig{fig:sps7_spectra}~a
to larger values.
For example, running the \ilc\ at the `canonical' energy $\sqrt{s}=500\,\GeV$ 
reduces the $\stau_1\stau_1$ rate by 10\,\%
and the sample of trapped $\stau's$ by a factor of two, 
degrading the precision on the $\stau$ lifetime to 2\,\%.

{\it Note:} 
The relative precision on the $\stau$ lifetime measurement does 
not depend on the gravitino mass, should it be much lighter as often assumed
in gauge mediated supersymmetry models. 
Technically, there may be a limitation 
to measure very short lifetimes below a millisecond.
In the present method the $\stau$ decay is required to occur 
uncorrelated to any beam collision, 
which essentially means outside a time interval of $\Delta t = 1\,\msek$
for a whole bunch train, repeating at a rate of 5\,Hz.
As an illustration,
in the SPS~7 scenario a $\stau$ lifetime of $5\,\msek$ corresponds to a
gravitino mass of $0.5\,\MeV$.

\subsection{Gaugino mediated symmetry breaking scenarios P3}

The model P3 is investigated at $\sqrt{s}=500\,\GeV$ and assuming 
$\cL=100\,\fbi$.
The inclusive $\stau$ cross section of $\sigma_{SUSY}=470\,\fb$
provides large rates, as can be seen in the $\beta\gamma$ spectrum of
\fig{fig:p13_spectra}~a.
The main contributions to the sample of stopped $\stau's$ come from
$\nt_1\nt_1$ and $\ser\sel$ production, one expects about 3900 and 3700
particles in the calorimeter and yoke, respectively.
From the kinematics of $\stau_1\stau_1$ production one gets
a $\stau$ mass of $m_{\stau_1}=102.5\pm 0.2\,\GeV$.
A fit to the $\stau$ lifetime distribution yields
$t_\stau=(4.2\pm 0.1)\cdot 10^7\,\sek$.
A direct measurement of the gravitino mass using
the $\tau$ recoil energy spectrum of $\stau\to\tau\sG$,
as displayed in \fig{fig:p13_spectra}~b,
provides a value of $m_\sG=25\pm 1.5\,\GeV$.
Alternatively,
one derives a gravitino mass of $m_\sG=25\pm 0.3\,\GeV$ from
the $\stau$ lifetime measurement.

{\it Note:}  The number of trapped $\stau's$ in the calorimeter
 can be increased by a factor of 1.8 when running at a slightly lower
 cms energy $\sqrt{s}=480\,\GeV$.

\begin{figure}[htb]
  \setlength{\unitlength}{1mm} 
  \begin{picture}(150,70)(0,0) 
    \put(75,0){
      \put(0,0){\epsfig{file=p3_fitspectra.eps,angle =90,width=80mm}} 
      \put(68,49){ \large\bf b) }      
      \put(12,64){\epsfig{figure=box.eps,height=5mm,width=50mm}}
      \put(50,58){ $m_\stau=103\,\GeV $}      
      \put(50,55){ $m_\sG=~25\,\GeV $}      
      \put(45,0){\epsfig{figure=box.eps,height=6mm,width=50mm}}
      \put(45,0){\large $\tau$ jet $E_{jet}~[\GeV] $}  }    
    \put(-15,0){
      \put(0,0){\epsfig{file=p3_betagamma.eps,angle=90,width=80mm}}
      \put(68,49){ \large\bf a) }      
      \put(52,58){{\color{red}\large \bf\boldmath $\cx$MSB P3}}
      \put(20,0){\large
        \put(-7,50){\epsfig{figure=box.eps,height=16mm,width=20mm}}
        \put(0,55){\color{magenta} $\stau_1\stau_1$}
        \put(0,50){\color{cyan} $\ser\se_{R,L}$}
        \put(0,45){\color{yellow} $\smur\smur$}
        \put(0,40){\color{green} $\nt_1\nt_1$} }
      \put(50,0){\epsfig{figure=box.eps,height=6mm,width=40mm}}
      \put(50,0){\large $p_\stau/m_\stau = \beta\gamma$}  }
  \end{picture}
  \caption{ a) Scaled momentum spectrum 
    $p/m = \beta\gamma$ of $\stau$ production
    with contributions from $\stau_1\stau_1$ (magenta), $\ser\se_{R,L}$ (cyan),
    $\smur\smur$ (yellow) and $\nt_1\nt_1$ (green);
     b) $\tau$ recoil spectra of the decay $\stau_1\to\tau\sG$ with
    $E_{jet}$ of $\tau\to\rho\nu_\tau + 3\pi\nu_\tau$ compared 
    with simulations assuming $m_\sG=25\,\GeV$ (red histogram), 
    $20\,\GeV$ (blue) and $30\,\GeV$ (magenta).
    $\cx$MSB~P3 scenario, $\cL=100\,\fbi$ at $\sqrt{s}=500\,\GeV$}
  \label{fig:p13_spectra}
\end{figure}

The scenario P3 yields a lifetime of about 1.3 years. 
Taking instead a larger gravitino mass of $m_\sG=50\,\GeV$, 
as proposed by the authors of \cite{bks}, the lifetime increases
by almost an order of magnitude, and it may be difficult to observe
all accumulated $\stau's$ during the operation of an experiment.
Assuming a measurement period of three years will provide a lifetime
of $t_\stau=(3.9\pm 0.7)\cdot 10^8\,\sek$, corresponding to 12.4 years.
The relative error grows faster than expected from the reduced statistics
(factor 4 less decays) when observing only a fraction of the lifetime.
Of coarse, this has also consequences for the measurement of the decay spectra.

The {\it $\cx$MSB~P1 scenario} is characterised by somewhat heavier 
sparticles and
is investigated at $\sqrt{s}=420\,\GeV$, just above threshold production of
sleptons and the lightest neutralino. 
Again, the event statistics allow measurements to be performed at
the per mill to per cent level, see \tab{tab:results}.

\section{Discussion of results}
\label{discussion}

If kinematically accessible, 
metastable staus are copiously produced at the \ilc.
They are easy to detect
and their properties, like mass and lifetime,
and the decays into gravitinos
can be studied in great detail. 
The results of the various case studies of the supersymmetry models
and the assumed experimental conditions
are summarised in \tab{tab:results}.
The common features of all spectra concerning the achievable accuracies 
of the basic measurements are:
(i)
the {\it $\stau$ mass} can be measured at the per~mill level
by exploiting the 2-body kinematics of $\stau_1\stau_1$ pair production
in the tracking detector;
(ii)
the {\it $\stau$ lifetime} can be determined within a few per~cent
by observing the decays $\stau\to\tau\sG$ triggered in the calorimeter
and/or iron yoke;
and 
(iii)
the {\it $\sG$ mass} can be accessed directly  with a precision
of order 10\,\% 
(degrading rapidly for $m_\sG/m_\stau \lesssim 0.1$) 
by measuring the $\tau$ recoil spectra of the $\stau$ decays 
in the calorimeter.

\renewcommand{\arraystretch}{1.1}
\begin{table}[htb] \centering
  \caption{Expected accuracies on the determination of $\stau$ and $\sG$
    properties for various \susy\ scenarios:
    stau mass $m_{\stau}$, stau lifetime $t_{\stau}$, 
    gravitino mass $m_{\sG}^{E_{\tau}}$
    from $\tau$ recoil energy spectra
    and
    gravitino mass $m_{\sG}^{t_{\stau}}$ deduced from $\stau$ lifetime.
    Experimental conditions 
    in the last columns: 
    centre of mass energy $\sqrt{s}$, 
    inclusive $\stau$ production cross section $\sigma_{\stau\stau X}$,
    integrated luminosity $\cL$, number of
    trapped $\stau's$ in hadron calorimeter $N_\stau^{\rm hcal}$
    and instrumented yoke $N_\stau^{\rm yoke}$  }  
  \label{tab:results}
  \vspace{2mm}
  \begin{tabular}{l c c c c c c c c c}
           &  $m_{\stau}\ [\GeV]$ 
           & $t_{\stau} \, [\sek]$
           & $m_{\sG}^{E_\tau} \, [\GeV]$
           & $m_{\sG}^{t_{\stau}} \, [\GeV]$  
           & ~ $\sqrt{s} \, [\GeV]$ 
           & $\sigma_{\stau\stau X} \, [\fb]$
           & $\cL \, [\fbi]$
           & $N_\stau^{\rm hcal}$ & $N_\stau^{\rm yoke}$
    \\[.5ex] \hdick 1
    SPS 7          & $ 124.3 \pm 0.1 $ & $ 209.3\pm 2.4 $
                   & $ < 9$  & $ 0.1 \pm 0.001 $ 
                   & $ 410 $ & $ 420 $ & $ 100 $ & $ 10000$ & $ 4900 $ \\ 2
    FS 600         & $ 219.3 \pm 0.2 $ & $ (3.6\pm 0.1)\,10^6 $
                   & $ 50 \pm 9 $      & $  50   \pm 0.7 $ 
                   & $ 500 $ & $  20 $ & $ 250 $ & $ 2100 $ & $ 4200 $ \\ 3
    GDM $\epsilon$ & $ 157.6 \pm 0.2 $ & \ $ (2.6\pm0.05)\,10^6 $ \
                   & $ 20 \pm 4$       & $  20   \pm 0.2 $ 
                   & $ 500 $ & $ 300 $ & $ 100 $ & $ 4100 $ & $ 1850 $ \\ 4
    GDM $\zeta$    & $ 340.2 \pm 0.2 $ & $ (1.8\pm 0.06)\,10^6 $
                   & $ 100 \pm 10 $  & $ 100   \pm 2  $ 
                   & $ 800 $ & $   5 $ & $ 1000$ & $ 1350 $ & $  800 $ \\ 5
    GDM $\eta$     & $ 322.1 \pm 0.2 $ & $ (6.9\pm 0.3)\,10^4 $
                   & $  20 \pm 25 $    & $  20   \pm 0.4 $ 
                   & $ 800 $ & $   9 $ & $ 1000$ & $ 1050 $ & $ 1850 $\\ 6
    $\cx$MSB P1 ~  & $ 185.2 \pm 0.1 $ & $ (9.1\pm 0.2)\,10^6 $
                   & $  50 \pm 3 $     & $  50   \pm 0.6 $
                   & $ 420 $ & $  27 $ & $ 200 $ & $ 3700 $ & $ 4100 $ \\ 7
    $\cx$MSB P3 ~  & $ 102.5 \pm 0.2 $ & $ (4.2\pm 0.1)\,10^7 $
                   & $  25 \pm 1.5 $   & $  25 \pm 0.3 $ 
                   & $ 500 $ & $ 470 $ & $ 100 $ & $ 3900 $ & $ 3700 $ \\
  \end{tabular}  
\end{table}

These expectations are based on moderate integrated luminosities, 
the event rates to be accumulated during one to three years of data taking 
at the \ilc\ under nominal conditions.
Although the $\stau$ decays are treated in a parametrised form the results
should be reliable within a factor of two. 
A detailed detector simulation may lead to a reduction of the data samples
due to acceptance cuts, quality criteria and resolution effects.
However, the losses may be compensated and
the event statistics can be considerably increased by 
optimising the beam energies (see sect.~\ref{analyses})
and making use of the beam polarisations.
For instance, the production cross sections for light sleptons,
$\stau_1\stau_1$, $\ser\ser$ and $\smur\smur$,
are enhanced by a factor $(1+\cP_{e^-})(1-\cP_{e^+})\simeq 2.9$
when using right-handed polarised electrons of degree $\cP_{e^-}= +0.8$ 
and left-handed polarised positrons of $\cP_{e^+}= -0.6$.

Concerning the expected \ilc\ detector performance with respect to 
detecting metastable $\stau's$ there is, however, a caveat. 
At present, it is envisaged to operate the LDC calorimeters
in a pulsed mode~\cite{ldc}. 
In order to cope with the heat production of the electronics, 
it is foreseen that the amplifiers and read out channels of the detector
parts will be switched on and read out just during one bunch train for 2~ms, 
and then be switched off to wait for the next bunch train occurring 
200~ms later; {\it i.e.} the detector is inactive for most of the time.
Such an operation is designed for usual, beam correlated physics scenarios,
but clearly has to be revised if one wants to observe 
the decays of metastable particles.
For the iron yoke instrumentation it does not really  seem to be necessary 
to apply a power pulsing.
A quasi-continuous read-out of all channels above a certain threshold 
over the so far 'idle' time gap of 198~ms appears possible,
replacing the $\stau$ decay trigger by software analysis~\cite{eckerlin}.
Thus a lifetime measurement will be feasible in any case, may be at the
expense of a lower event rate restricted to $\stau's$ trapped in the yoke.
The goal to also increase the duty cycle of the hadron calorimeter
substantially has to be pursued in future R\&D detector developments.
Since a direct measurement of the gravitino mass is extremely important,
it will be assumed in the following discussion that the technical 
problems will be solved and the calorimeter will be permanently sensitive
with essentially no dead-time.

An independent measurement of the gravitino mass is of prime
importance. 
Applying \eq{eq:planckscale} one can determine the supergravity Planck scale
from the NLSP decay with an accuracy of order 10\'\%. 
A value consistent with the macroscopic scale of gravity,
{\it i.e.} Newton's constant $G_N=6.7\cdot10^{-39}\,\GeV^{-2}$, would be a
decisive test that the gravitino is indeed the superpartner of the graviton.
At the same time the measurement of the gravitino mass can be used to yield
the supersymmetry breaking scale $F=\sqrt{3}\,M_P\,m_{3/2}$, 
see \eq{eq:gravitinomass},
which is an important parameter to unravel the nature of the supersymmetry 
breaking mechanism.

From the $\stau$ mass and lifetime measurements one can compute the gravitino
mass  {\it assuming} that the $\stau$
decay occurs with gravitational coupling according
to \eq{eq:decaywidth}. The expected accuracy is again of order per~cent.
This interpretation may be plausible, however,
it can only be justified, if there is additional
information which confirms the gravitino LSP assignment.  

For too light gravitinos a direct mass measurement may not be possible, leaving
the nature of the LSP open. 
An alternative option may be a spin $1/2$ axino $\sa$ as lightest
supersymmetric particle~\cite{brandenburg}. 
In such a scenario
the $\stau$ lifetime is not related to the axino mass and,
depending on the parameters, 
may vary between some 0.01\,sec and 10\,hours, 
comparable to the expectations of gauge mediated and gravity mediated \susy\
breaking models.
A method to differentiate between these two scenarios is to study the
radiative three-body decays $\stau\to\tau\gamma\sG$ and
$\stau\to\tau\gamma\sa$. 
The branching ratios are suppressed by two orders of magnitude and
expected to be slightly larger for the axino LSP, the exact values depend on
the selected phase space to identify an isolated photon.
More promising is to investigate $\gamma-\tau$
correlations of the decay topology.
In the case of gravitino LSP the photons are preferentially emitted 
collinear with the $\tau$ and have a soft, Bremsstrahlung like energy spectrum.
A similar configuration also occurs for the axino LSP, however, there is in
addition a substantial rate of energetic photons,
$E_\gamma \to m_\stau/2$, emitted opposite, back-to-back
to the $\tau$ direction.
Such a signature would clearly allow to distinguish between the interpretations
as axino LSP or as gravitino LSP.
Experimentally the analysis of radiative $\stau$ decays
is quite ambitious because one has to 
discriminate a single photon against the photons from $\pi^0$ decays 
as well as against the hadrons in hadronic $\tau$ decays. 
The expected low event rates will be further reduced by
efficient selection criteria.
Hence, large statistics data samples are required, which may be provided
by the \ilc.

\section{Conclusions}

It has been shown in detailed analyses
that future \ilc\ experiments have a rich potential to 
study \susy\ scenarios
where the gravitino $\sG$ is the lightest supersymmetric particle 
and a charged stau $\stau$ is the long-lived, metastable 
next-to-lightest supersymmetric particle.
Precise determinations of the $\stau$ mass and lifetime 
and of the $\sG$ mass appear feasible, 
provided the proposed detectors will operate at a reasonable duty cycle.
A calorimetric measurement of the gravitino mass 
from the $\tau$ recoil spectra of the  decay $\stau\to\tau\sG$ 
gives access to the gravitational coupling, 
{\it i.e.} to the Planck scale, in a microscopic particle experiment
and thus provides a unique test of supergravity.
Furthermore these observations will put stringent constraints on 
an interpretation of the gravitino as dark matter candidate,
being undetectable in astrophysical search experiments,
and will allow its relic density to be to computed reliably.

\bigskip\noindent
{\it Acknowledgements.} \ 
I want to thank W. Buchm\"uller and P. M. Zerwas for many valuable discussions
and comments on the manuscript.


%

\end{document}